\documentclass[aps, prd, groupedaddress, preprint, tightenlines,  eqsecnum, nofootinbib, showpacs]{revtex4}
\usepackage{epsfig}

\usepackage{axodraw, float, slashed, graphicx, amssymb, amsmath}
\usepackage[usenames, dvipsnames]{xcolor}
\usepackage{booktabs}

\usepackage[margin=2.2cm, a4paper, includefoot]{geometry}

\def\npb#1#2#3{{\rm Nucl. Phys. B}{\bf \ #1}, #3 (#2)}
\def\spa#1.#2{\left\langle#1\,#2\right\rangle}
\def\spb#1.#2{\left[#1\,#2\right]}
\def\eps{\epsilon}
\def\la{\langle}
\def\ra{\rangle}

\def\Aloop{A^{\oneloop}}
\def\BR#1#2{[#1|{K_{abc}}|#2\ra}
\def\BRTTT#1#2{\la#1^+|\slashed{K}_{abc}|#2^+\ra}
\def\NeqEight{\Neq8}
\def\NeqSix{\Neq6}
\def\NeqFour{\Neq4}
\def\NeqOne{\Neq1}

\def\deff{d_{\text{eff}}}

\def\coli#1#2{\mathop{\longrightarrow}^{#1 \parallel #2}}

\newcommand{\oneloop}{\text{1-loop}}

\newcommand{\tree}{\text{tree}}
\newcommand{\Neq}[1]{\mathcal{N} = #1}

\newcommand\trunc{\text{trunc}}
\DeclareMathOperator{\Perm}{\mathcal{P}}

\DeclareMathOperator{\tr}{\mathrm{tr}}
\DeclareMathOperator{\SP}{\mathrm{Sp}}
\DeclareMathOperator{\Soft}{\mathrm{Soft}}

\def\SoftP{\hat S}

\def\be{\begin{equation}}
\def\ee{\end{equation}}

\def\captionfigA{The box and bubble functions appearing in the $\NeqFour$ MHV
      one-loop amplitude}

\def\captionfigD{The loop part of a link diagram producing a term in
  $R_n^{r}$.
The  $r$ positive helicity legs of $P^r$ lie in the loop.}

\def\captionfigE{A link diagram corresponding to a term in 
$C_{n-10}  \times  \SoftP^{8}[P^{n-10};Q^8]$}

\def\captionfigG{The topologies of the link diagrams for the seven
  point MHV amplitude. The vertices are labelled by the five positive
  helicity legs. 
}
\def\captionfigK{The $K$-vertex in a link diagram for the $n-1$ point
  amplitude. The vertex may or may not be part of the loop.}

\def\captionfigL{The corresponding link diagrams in the $n$-point amplitude} 

\def\captionfigM{These link diagrams do not contribute to $R^2_{n-1}$ in the $a^+b^+$
  collinear limit but cancel against terms arising from the box
  integral contribution.} 

\def\captionfigN{The diagrams contributing to the right-hand diagram
  in the soft limit of leg $n$.} 
\def\captionfigO{Diagrammatic representation of the soft-lifting
  functions. The particular figure contributes to $\SoftP^8[P^4;Q^8]$} 
\begin{document}

\title{Constructing Gravity Amplitudes from Real Soft and Collinear Factorisation}

\author{David~C.~Dunbar, James~H.~Ettle and Warren~B.~Perkins}

\affiliation{
College of Science, \\
Swansea University, \\
Swansea, SA2 8PP, UK\\
%\today
}

\begin{abstract}

Soft and collinear factorisations can be used to construct
expressions for amplitudes in theories of gravity. 
We generalise the ``half-soft'' functions used previously to
``soft-lifting'' functions and  use these to generate 
tree and one-loop amplitudes. In particular we 
construct expressions for MHV tree amplitudes and
the  rational terms in one-loop amplitudes in
the specific context of $\NeqFour$ supergravity.  
To completely determine the rational terms collinear
factorisation must also be used.   
The rational terms for $\NeqFour$ have a remarkable
diagrammatic interpretation as
arising from algebraic link diagrams.

\end{abstract}

\maketitle

\section{Introduction}

The $S$-matrix of a weakly coupled  Quantum Field Theory is a key
object which largely defines the theory and its interactions.  The
Feynman diagram approach is a very robust general method which, in
principle, may be used to compute the $S$-matrix.  Unfortunately this
method is very complex, particularly in theories which contain large
symmetries such as gauge theories and
theories of gravity. Computing the $S$-matrix from the constraints it must
satisfy is an old idea~\cite{Eden} which has seen huge development in
recent years using ideas based on
unitarity~\cite{Bern:1994zx,Bern:1994cg,Bern:1997sc,Britto:2004nc} 
and the factorisation properties of amplitudes~\cite{Britto:2005fq,BernChalmers}.

In this article, we examine soft-factorisation on real
kinematics in theories of gravity and  introduce  ``soft-lifting'' functions  
which allow us to express the 
$n$-graviton  ``Maximally-Helicity-Violating'' (MHV) tree amplitudes as a
``soft-lift'' of either the three or four-point tree amplitudes.  The
soft-lift of the three point amplitudes gives an expression
equivalent to previous forms but the soft-lift of the four-point
amplitudes yields a novel expression for  the MHV tree amplitude.

The same
soft-lifting functions are key elements in the
rational terms of $\NeqFour$ supergravity one-loop MHV amplitudes.
In this one-loop example real soft-limits  must be
combined with information from the collinear limits to obtain the
$n$-point expression.  
The expression for the  $n$-point rational term
was proposed in ref.~\cite{Dunbar:2011dw}
where numerical checks were applied to it. 
Here we show that these rational terms have a remarkably
simple interpretation in terms of one-loop link diagrams in the same
way as has been found for the MHV tree amplitude~\cite{Nguyen:2009jk}.  
Using the diagrammatic representation,  
we present an analytic
proof that the rational expressions have the correct soft and collinear
limits.

\vskip 1.0 truecm

\section{MHV tree amplitudes,  Half-Soft Functions and Twistor Link Diagrams}

MHV tree amplitudes for graviton scattering have been presented in a
wide variety of forms. In this section we review a range of these
and the relations between them. 
The original 
Berends, Giele and Kuijf (BGK)
form of the MHV gravity amplitude~\cite{BerGiKu} is\footnote{
The normalisation of the physical amplitude 
${\cal  M}^\tree=i(\kappa/2)^{n-2} M^\tree, 
{\cal  M}^\oneloop=i(\kappa/2)^{n} M^\oneloop$. As usual we are using a spinor helicity formalism with the usual
spinor products $ \spa{j}.{l} \equiv \langle j^- | l^+ \rangle =
\bar{u}_-(k_j) u_+(k_l)$ and $\spb{j}.{l}\equiv \langle j^+ | l^-
\rangle = \bar{u}_+(k_j) u_-(k_l)$, and where $\BR{i}{j}$ denotes
$\BRTTT{i}{j}$ with $K_{abc}^\mu =k_a^\mu+k_b^\mu+k_c^\mu$ etc. Also
$s_{ab}=(k_a+k_b)^2$, $t_{abc}=(k_a+k_b+k_c)^2$, etc.}
\begin{align}
M_n^{\rm tree}
&(1^-,2^-,3^+,\cdots, n^+) \;=\;(-1)^n \spa1.2^8 \times
\notag \\
& 
\biggl[{ \spb1.2\spb{n-2}.{n-1}    \over \spa1.{n-1} \prod_{i=1}^{n-1} \prod_{j=i+1}^n
  \spa{i}.j
 }
\Bigl(  \prod_{i=1}^{n-3} \prod_{j=i+2}^{n-1} \spa{i}.j \Bigr)
\prod_{l=3}^{n-3} ([n|K_{l+1\ldots n-1}|l\ra)
% \notag \\&\hspace{8.7cm}
+{\cal P}_{(2,3,\cdots,n-2)}
\biggr]\,.
\label{BGKform}
\end{align}
where ${\cal P}_{(2,3,\cdots,n-2)}$ indicates summing over the $(n-3)!$
permutations of legs $2,\ldots, n-2$.
The tree amplitude has the, non-manifest,  symmetry property that $M_n^{\rm tree}
(1^-,2^-,3^+,\cdots, n^+)/\spa1.2^8$ is crossing symmetric under all
exchanges of legs including the two negative helicity legs.  This fact may be proven
(or is due to) by recognising the $n$-graviton tree amplitude is the
same in pure gravity and $\NeqEight$ supergravity  and then looking at
the implications of the supersymmetric Ward identities.  The argument
for $\NeqEight$ supergravity follows from that for supersymmetric
Yang-Mills~\cite{SWI}  using the $\NeqEight$ Ward identities given,
for example, in appendix~E of ref.~\cite{Bern:1998ug}.
The BGK formulae was established using the Kawai, Llewellen and Tye
(KLT) 
relations~\cite{Kawai:1985xq}  which relate
gravity amplitudes to products of Yang-Mills amplitudes for lower
point functions and then by
verifying its soft-factorisation
properties~\cite{BerGiKu}. Subsequently alternative proofs of the
formulae have been presented~\cite{Mason:2008jy}.    MHV amplitudes
are an important  component of gravity theories and can be promoted to
fundamental vertices to construct other tree
amplitudes~\cite{BjerrumBohr:2005jr} as in Yang-Mills theories~\cite{Cachazo:2004kj}.

Gravity amplitudes have soft-limit singularities~\cite{BerGiKu}
as $k_n \longrightarrow 0$,
\be
M_n( \cdots , n-1, n^h) \longrightarrow \Soft_{n^h} \times M_{n-1} (\cdots ,
n-1 ) 
\ee
where the positive helicity ``soft factor'' is given by
\be
\Soft_{n^+}=-{ 1\over \spa{a}.n\spa{n}.{b} } \sum_{j\neq n,a,b} {
\spb{j}.n \spa{a}.j  \spa{j}.{b} \over \spa{j}.n } .  
\label{SoftFactorEq}
\ee
with $\Soft_{n^-}$ given by conjugation. 
The soft factor is independent of the choice of $a$ and $b$ ($\neq n$)
although this is not manifest. Recently it has also been suggested~\cite{ArkaniHamed:2009dn} that soft limits may
be used to determine an $n$-point amplitude
in terms of  $n-1$ point amplitudes multiplied by a
soft factor. This process, referred to as ``inverse-soft'',  has been
applied to gravity tree amplitudes~\cite{Nguyen:2009jk,BoucherVeronneau:2011nm,Bullimore:2010pa}
and used, for example,  to determine the MHV
amplitudes in the form,
\be
M_n^{MHV}  (1,2,\cdots  n)
=\sum_{i=1}^{n-2}
{\cal G}  (n-1,n, \hat i )
\times 
M_{n-1}^{MHV}(1,\cdots \hat i , \cdots \widehat{n-1} )
\label{InverseSoftEX}
\ee
where ${\cal G}$ denotes the soft-factor 
\be
{\cal G} (a,b,i) =-{ \spa{a}.i^2 \over \spa{a}.b^2 } {\spb{b}.i \over
  \spa{b}.i }
\ee
In this expression the $n-1$-point amplitude is evaluated at a complex
kinematic point where the legs $i$ and $n-1$ have been shifted:
\begin{align}
\hat k_i&= \lambda_i(
\bar \lambda_i  +  { \spa{n}.{n-1} \over \spa{i}.{n-1} }\bar\lambda_n
) \; , 
%\notag \\
\;\;\;\;\;
\hat k_{n-1}
%&
= \lambda_{n-1} (
\bar \lambda_{n-1}  +  { \spa{n}.i \over \spa{n-1}.i }\bar\lambda_n   ).
\label{SoftShiftEQ}
\end{align}
These complex momenta satisfy
\be
\hat k_i +\hat k_{n-1} = k_i+k_{n-1}+k_n
\ee
Although this expression can be regarded as an inverse-soft relation it is very
closely related to BCFW recursive expressions for gravity
\cite{Britto:2005fq,Bedford:2004nh}.   Other related expressions for the MHV
tree amplitude exist~\cite{Hodges:2011wm,Heckman:2011qu}.

Another  representation of the MHV tree amplitudes was given in~\cite{MaxCalcsB} in terms of  ``half-soft'' functions.  These
originally appeared in the box-coefficients of the one-loop MHV amplitude in
$\NeqEight$ supergravity and can be thought of as an off-shell
version of the tree amplitude. 
The half-soft
functions have the explicit form
\begin{align}
h(a,\{1,2,\ldots,m\},b) &\equiv \frac{\spb1.2}{\spa1.2} { [3
    | {K_{12}} |a \ra [4 | {K_{123}} |a \ra \cdots [m| {K_{1\cdots
        m-1}} |a \ra \over \spa2.3\spa3.4 \cdots \spa{m-1,}.{m} \,
    \spa{a}.1 \spa{a}.2\spa{a}.3 \cdots \spa{a}.{m} \, \spa1.{b}
    \spa{m}.{b} } 
\notag \\  &\hskip1cm + \Perm_{(2,3,\ldots,m)}, 
\label{NonRecursiveH}
\end{align}
When the half-soft functions are at the maximum size for an $n$-point
amplitude we have 
\be
M_n^{\rm tree}
(1^-,2^-,3^+,\cdots, n^+)
=   (-1)^n \spa{1}.{2}^6 \times h(1,\{3,4,\cdots, n\},2)
\label{SoftisTree}
\ee
Note that this implies $h(1,\{3,4\cdots, n\},2)/\spa1.2^2$ is
completely crossing symmetric.

An alternative expression for the half-soft functions was also given
in~\cite{MaxCalcsB} 
\be
h(a,\{1,2,\cdots m\},b) =\sum_{i_1,i_2,\cdots i_m=0}^{m-2}
\phi_m (i_1,i_2,i_3,\cdots i_m )\times \prod_{j=1}^m  (\spa{a}.j\spa{j}.b)^{i_j-1}
\label{halfsoftalt}
\ee
where the sum is restricted to $\sum_j i_j =m-2$. 
The $\phi_m$ are  polynomial in the objects
\be
\hat A [a;b] ={\spb{a}.b \over \spa{a}.b }
\ee
and are symmetric functions of their arguments. They 
are given recursively by
\be
\phi_m(i_1,i_2,i_3,\cdots 0 ) = \sum_{j=1}^{m-1} \phi_{m-1}(
i_1,i_2,\cdots  i_j-1, \cdots i_{m-1} ) \times 
\hat A [j; m] 
\ee
with the initial value
\be
\phi_2(0,0)\equiv \hat A[1;2] =  {\spb1.2\over\spa1.2}
\ee
In ref.~\cite{MaxCalcsB} an interpretation of the $\phi_m$ was given
in terms of Young Tableaux. 

In a more recent
development~\cite{Nguyen:2009jk}  an equivalent expression for the
tree amplitude was given
where the terms were interpreted as arising from tree ``link diagrams''
\be
M_n^{MHV} 
={(-1)^n \spa1.2^6 }
\sum_{\rm trees} \left(
\prod_{{\rm edges:}\, ab} {
\spb{a}.b \over \spa{a}.b }
\right)
\left(
\prod_{{\rm vertices:}\, a}  
( \spa{a}.1\spa{a}.2 )^{deg(a)-2} 
\right)
\label{TreeLinkForm}
\ee
The tree link diagrams are all the connected graphs which can be drawn
between $n-2$ labelled vertices representing the positive helicity legs. Vertices with any number of legs (or
degree $\deg(a)$) are allowed.  For example, for the seven point
amplitude we have the 125 diagrams obtained by the labellings of the
topologies given in figure~\ref{TreeFigure}.

\begin{figure}[H]
  \begin{center}
    {
      \begin{picture}(150,120)(0,0)
    \SetOffset(50,60)
\SetWidth{2}
\Line(-40,0)(80,0)
\SetWidth{1}
\BCirc(-40,0){5}
\BCirc(-10,0){5}
\BCirc(20,0){5}
\BCirc(50,0){5}
\BCirc(80,0){5}
     \end{picture}
    \begin{picture}(150,120)(0,0)
    \SetOffset(25,60)
\SetWidth{2}
\Line(20,0)(80,0)
\Line(20,0)(0,20)
\Line(20,0)(0,-20)
\SetWidth{1}
\BCirc(0,20){5}
\BCirc(0,-20){5}
\BCirc(20,0){5}
\BCirc(50,0){5}
\BCirc(80,0){5}
     \end{picture}
    \begin{picture}(150,120)(0,0)
    \SetOffset(0,60)
\SetWidth{2}
\Line(20,0)(80,0)
\Line(50,0)(50,30)
\Line(50,0)(50,-30)
\SetWidth{1}
\BCirc(50,30){5}
\BCirc(50,-30){5}
\BCirc(20,0){5}
\BCirc(50,0){5}
\BCirc(80,0){5}
     \end{picture}
     }
    \caption{\captionfigG \label{TreeFigure} }
  \end{center}
\end{figure}

There is an obvious link diagram interpretation of the half-soft functions of maximum size following from~(\ref{SoftisTree}) 
and~(\ref{TreeLinkForm}). From~(\ref{halfsoftalt}) it can be seen that the  
same rules also generate the  half-soft functions with restricted sets to positive helicity legs.

In the next section,  we  introduce ``soft-lifting'' functions which are
generalisations of the half-soft functions. 
In some regards, we are following the
spirit of ``inverse soft'' however we will be using real momenta. 
We apply these to
evaluate  MHV tree amplitudes and  the rational parts of one-loop MHV amplitudes
in $\NeqFour$ supergravity.

We find that we need to consider both collinear and soft limits in order to determine one-loop amplitudes.    
Gravity amplitudes are not
singular in the collinear limit $k_a\cdot k_b\longrightarrow 0$,  but
acquire collinear phase singularities which take a form that is specified in
terms of amplitudes with one fewer external leg~\cite{MaxCalcsB}. Specifically, if
$k_a \longrightarrow z K$ and $k_b \longrightarrow (1-z) K $,
\be
M_n( \cdots , a^{h_a}, b^{h_b} ) \coli{a}{b}   \sum_{h'}
\SP_{-h'}^{h_ah_b}  M_{n-1} (\cdots , K^{h'} )   +F_n
\ee
where the $h$'s denote the various helicities of the
gravitons and $F_n$ is free of phase singularities.
The 
``splitting functions'' are~\cite{MaxCalcsB}
\begin{align}
  \SP_{+}^{-+} 
= -{ z^3\spb{a}.{b} \over (1-z) \spa{a}.{b} },
\;\;\; 
  \SP_{-}^{++}=-{ \spb{a}.{b} \over z(1-z)  \spa{a}.{b}  },  
\;\;\; 
  \SP_{+}^{++}=0  .
\end{align} 
with the others obtained by conjugation.

\section{Soft-Lifting Functions}
\def\m{p} 
Analyses of soft and collinear divergences have long been used as
tools for constructing amplitudes~\cite{BerGiKu} and indeed there are
recent suggestions that gravity scattering amplitudes can be
determined from their soft-behaviour alone~\cite{ArkaniHamed:2009dn}.  
In attempting to do this it is useful to define building blocks which have simple soft and
collinear behaviour.  For example, the   ``half-soft''
functions of ref.~\cite{Bern:1998ug} were used by the authors to
construct the all-plus graviton one-loop amplitude.  Here we
generalise the half-soft functions to ``Soft-Lifting Functions'' which
we will use in several constructions.

We define  soft-lifting functions, $\SoftP^\m[ P^s ;Q^\m ]$, where $P^s=\{p_j\}$ and $Q^\m=\{q_k\}$
are disjoint sets of the positive helicity legs of length $s$ and $\m$
respectively.
When the set $P^s$ is of length
$1$ the soft-lifting functions are the half-soft
functions of ref.~\cite{MaxCalcsB}  up to a factor,
\be
\SoftP^{\m}[ \{p_1\}; Q^\m] \equiv ( \spa{m_1}.{p_1}^2\spa{m_2}.{p_1}^2) \times  h(m_1,Q^\m +p_1,m_2)
\label{SoftOneEQ}
\ee
From this we define the general soft-lifting function by 
\be
\SoftP^{\m}[ P^s; Q^\m] =
 \sum_{\rm partitions:\beta} \prod_{j=1}^s \SoftP^{\m}[ \{p_j\} ; Q_j] 
\label{SoftDefineEQ}
\ee
where the sum is over all partitions of the $q_k\in Q^\m$ into subsets
$Q_j$ where $Q_1 \cup Q_2 \cdots \cup Q_s=Q^\m$.  The summation
includes the terms where some of the $Q_j$ are null.  By definition, $\SoftP^0=1$.

We use shortened notation   $\SoftP^\m[ Q^\m]$ for  the special case of $\SoftP^\m[ P^s ;Q^\m ]$ when
$P^s=P^+-Q^\m$ where $P^+$ is the full set of positive helicity legs
(usually $\{ 3,4,\cdots n\} $).  In this case $s=n-2-p$. 
In this section we label the two
negative helicity legs  as $m_1$ and $m_2$
(typically these are the legs $1$ and $2$).

The soft-lifting functions are
polynomial in the ``soft-components'' $A[p_i;q_k]$ and $A[q_l;q_k]$ where
\be
A[p;q]  \equiv \SoftP^1[\{p\}; \{q\}] = { \spb{q}.{p}\spa{p}.{m_1}\spa{p}.{m_2}   
\over \spa{q}.{p}
  \spa{q}.{m_1}\spa{q}.{m_2} }  \;\;\;\;  q \neq m_1,m_2,p
\ee
In terms of the soft-components, 
\be
\SoftP^1[P^s ; \{q_1\} ] \equiv  \sum_{p_j\in P^s}   A[p_j;q_1]
\ee
Note that we could include $m_1$ and $m_2$ in the summation with no
change in value since $A[m_1;q]=A[m_2;q]=0$ so that 
$\SoftP^1[ \{q_1\} ]  = \sum_{j\neq q_1}   A[p_j;q_1]$.
The $\SoftP^\m$ are products of $\SoftP^1$ with cycle terms removed, 
\be
\SoftP^\m[P^s;Q^\m] =  \prod_{k=1}^\m  \SoftP^1[P^s+Q^\m-q_k;\{q_k\}]-\hbox{\rm cycle terms} . 
\label{CycleRemoveEQ}
\ee
where a cycle term is a  cyclic combination of
 $A[q_i; q_j]$,  that is terms of the form  $A[q_1;q_2]A[q_2;q_3]\cdots A[q_n;q_1]$.
The first few are given by,
\begin{align}
\SoftP^2[ P ; {\{q_1,q_2\}} ]&
= \SoftP^1[P\cup\{q_2\};{q_1}] \SoftP^1[P\cup\{q_1\};{q_2}] -A[q_1;q_2]A[q_2;q_1]
\notag \\ 
&
=
\sum_{p_j\in P}  A[p_j;q_1]
\sum_{p_k\in  P} A[p_k;q_2]
+ A[q_1;q_2]\sum_{p_j\in P}  A[p_j;q_1] 
+ A[q_2;q_1]\sum_{p_j\in P}  A[p_j;q_2]
\notag \\ 
\SoftP^3[ P;  {\{q_1,q_2,q_3\}} ]&
= \SoftP^1[P\cup\{q_2,q_3\};\{q_1\}] \SoftP^1[P\cup\{q_1,q_3\};\{q_2\}] \SoftP^1[P\cup\{q_1,q_2\};\{q_3\}] 
\notag \\ 
&\hskip 0.5 truecm  -\Biggl( A[q_1;q_2]A[q_2;q_1]\SoftP^1[P\cup\{q_1,q_2\};\{q_3\}]
+{\cal P}_{q_1q_2q_3} \Biggr)
\notag\\
&= \sum_{p_j\in P}  A[p_j;q_1]
\sum_{p_k\in  P} A[p_k;q_2]\sum_{p_l\in P}  A[p_l;q_3]
\notag\\
&+\biggl(
\sum_{p_j\in P}  A[p_j;q_1]
\sum_{p_k\in  P} A[p_k;q_2] ( A[q_1;q_3]+A[q_2;q_3]) +{\cal
  P}_{q_1q_2q_3}  \biggr)
\notag \\
+\biggl( 
\sum_{p_j\in P}  A[p_j;q_1] & \Bigl( A[q_1;q_2]A[q_1;q_3]+A[q_1;q_2]A[q_2;q_3]+A[q_3;q_2]A[q_1;q_3]\Bigr) +{\cal
  P}_{q_1q_2q_3}  \biggr)
\label{ExplicitEQ}
\end{align}

We can also construct a diagrammatic  representation of 
$\SoftP^\m[P^s;Q^\m]$.
This representation is based on  tree structures emanating
from ``seed points''  $p_j\in P^s$.   The rules for constructing these trees
are as follows: vertices $q_k$ must be attached to either the $p_j$
directly or to other $q_{k'}$ vertices.  The links are directional, with a link from
$q_k$ to $x$ giving a contribution of $A[x;q_k]$. The arrows on the links are necessary since $A[p_i;q_k]\neq A[q_k;p_i] $.   Closed loops are not
allowed.  Individual $p_j$ need not be connected to any $q_k$.  Each
$q_k$ has exactly one link starting from it, but may have any number
entering it.   An example graph contributing to $\SoftP^8[P^4;Q^8]$ is shown
in fig.~\ref{FigO}.    This representation differs from that for the
MHV tree and $R_n$ rational term but is closely related.  
Each $q_k$-dependent factor is $(\spa{1}.{q_k}\spa{2}.{q_k})^N$ where 
\be
N={\rm number\ of\; incoming\; arrows-number\; of\;
  outgoing\; arrows}  = {\rm deg(q_k)-2} 
\ee
since there is precisely one outgoing line.  However the $p_j$
dependent factor is just $(\spa{1}.{p_j}\spa{2}.{p_j})^{\rm deg(p_j)} $.  
So we can represent the soft-lifting function as
\begin{align}
\SoftP[P^s;Q^r]
&=
\sum_{\rm graphs} \left(
\prod_{{\rm edges:}\, a\rightarrow b} 
{  A[a;b]   }
\right)
\notag \\
&=
\sum_{\rm graphs} \left(
\prod_{{\rm edges:}\, ab} 
{\spb{a}.b \over \spa{a}.b }
\right)
\left(
\prod_{{\rm vertices:}\, q_k}  
( \spa{q_k}.1\spa{q_k}.2 )^{deg(q_k)-2} 
\right)
\left(
\prod_{{\rm vertices:}\, p_j}  
( \spa{p_j}.1\spa{p_j}.2 )^{deg(p_j)}  
\right)
\notag \\
&=
\left(
\prod_{p_j}  
( \spa{p_j}.1\spa{p_j}.2 )^{2}  
\right)
\sum_{\rm graphs} \left(
\prod_{{\rm edges:}\, ab} 
{\spb{a}.b \over \spa{a}.b }
\right)
\left(
\prod_{{\rm vertices:}\, a}  
( \spa{a}.1\spa{a}.2 )^{deg(a)-2} 
\right)
\label{twigs}
\end{align}
In the final form we have the vertex and link rules as for the MHV trees, but with a pre-factor and no connections between the $p_j$.

\begin{figure}[H]
  \begin{center}
    {
      \begin{picture}(450,120)(0,0)
   \SetOffset(50,20)
\SetWidth{2}
\ArrowLine(50,70)(50,0)
\ArrowLine(130,50)(130,0)
\ArrowLine(110,80)(130,50)
\ArrowLine(150,80)(130,50)
\ArrowLine(210,30)(230,0)
\ArrowLine(250,30)(230,0)
\ArrowLine(230,40)(230,0)
\ArrowLine(230,80)(230,40)
\BCirc(50,0){5}
\BCirc(130,0){5}
\BCirc(230,0){5}
\BCirc(310,0){5}
\BCirc(50,70){5}
\BCirc(130,50){5}
\BCirc(110,80){5}
\BCirc(150,80){5}
\BCirc(210,30){5}
\BCirc(250,30){5}
\BCirc(230,40){5}
\BCirc(230,80){5}
\Text(35,0)[]{$p_1$}
\Text(115,0)[]{$p_2$}
\Text(215,0)[]{$p_3$}
\Text(295,0)[]{$p_4$}
\Text(35,70)[]{$q_1$}
\Text(115,50)[]{$q_2$}
\Text(95,80)[]{$q_3$}
\Text(165,80)[]{$q_4$}
\Text(195,30)[]{$q_5$}
\Text(265,30)[]{$q_6$}
\Text(220,45)[]{$q_7$}
\Text(215,80)[]{$q_8$}
      \end{picture}
\\
\caption{\captionfigO\label{FigO} }
}
  \end{center}
\end{figure}

In general $\SoftP^\m[ P^s; Q^\m]$ satisfies the following iterative
definition: setting
\be
T^\m[P^s;Q^\m]\equiv
\prod_{k=1}^{\m} \hat S^1[ P^s; \{q_k\}] 
\ee
we have
\begin{align}
\SoftP^\m[ P^s; Q^\m ]
&= 
%T^\m [P^s, Q^\m]
%\notag \\&
\sum_{r=1}^{\m} \sum_{Q_1\subset Q^\m; |Q_1|=r}
T^{r}[ P^s; {Q_1} ] \times 
\SoftP^{\m-r}[ Q_1; {Q^\m-Q_1}]
\label{RecursiveSoft}
\end{align}
To see this from the diagrammatic representation,  consider the subset of
diagrams where each of the r-vertices of subset $Q_1\subset Q^\m$  are attached directly to the
$p_j\in P^s$. They can be attached to any of the $p_j$ and, summing over the possibilities, 
the links joining the $q_k$ to the $p_j$ give a factor of 
\be
\sum_{\rm partitions:\rho} \prod_{k=1}^r A[p(\rho,k),q_k]
\equiv\prod_{k=1}^r\sum_{p_j\in P^s} A[p_j;q_k] = T[P^s;Q_1]
\ee
where the sum is over all partitions of the $q_k\in Q_1$ amongst the $p_j\in P^s$ and $p(\rho,k)$ denotes which of the $p_j$ vertices 
the $q_k$ vertex is attached to in partition $\rho$. 
In this subset of diagrams the remaining $q\in Q^\m-Q_1$ vertices are not attached directly the $p_j$ vertices, so they must attach to the 
$q_k\in Q_1$ vertices in any combination. Using the diagrammatic representation these links yield a factor of
$\SoftP^{m-r}[Q_1;Q^m-Q_1]$. Overall this subset of diagrams yields
\be
T^r[P^s;Q_1] \times \SoftP^{m-r}[Q_1;Q^m-Q_1]
\ee
Summing over $r$ and all possible choices of $Q_1$ gives eq.~(\ref{RecursiveSoft}). 

We finish this section by noting the soft behaviour of the soft-lifting
functions, 
\begin{align}
 \SoftP^\m[P^s;Q^\m]
\longrightarrow&   -\Soft_{q_1^+} (m_1, P^s + Q^\m-q_1 , m_2 )
\times \SoftP^{\m-1}[P^s,Q^\m-q_1]
\notag\\
 \SoftP^\m[Q^\m]
\longrightarrow&   -\Soft_{q_1^+} \times \SoftP^{\m-1}[Q^\m-q_1]
\end{align}
for $q_1\in Q^\m$ with no soft-singularity if $q_1\notin Q^\m$.

\section{Relationship to MHV tree amplitudes}

Our first application  of the soft-lifting functions is to re-express the $n$-point
MHV amplitude as a ``soft-lift'' of the three-point tree amplitude
$M_3^{MHV}(1^-,2^-,p^+)$,  for any choice of positive helicity leg $p$,
\be
M_n^{MHV} = (-1)^{n-3}
\hat M_3^{MHV}(1^-,2^-,p^+) \times \SoftP^{n-3}[P^+-\{p\}]
\label{TreeLiftingEQ}
\ee
where
\be
\hat M_3^{MHV}(1^-,2^-,p^+)\equiv 
- {\spa1.2^6 \over \spa1.p^2\spa2.p^2 }
\ee
and $P^+=\{3,4,\cdots n\}$ is the set of all positive helicity legs. 
 This expression is
just a relabelling of
eq.~(\ref{SoftisTree}) using eq.(\ref{SoftOneEQ})  in a suggestive form.

We can also soft-lift the four-point amplitude to a new expression
for the $n$-point amplitude
\be
M_n^{MHV} =
{(-1)^{n-4}\over (n-3) } \sum_{\{p_1,p_2\}\subset P^+}    \hat M_4^{MHV}(1^-,2^-,p_1^+,p_2^+) \SoftP^{n-4}[ P^+-\{p_1,p_2\}]
\label{TreeLiftingEQ4}
\ee
where
\be
\hat M_4^{MHV}(1^-,2^-,p_1^+,p_2^+)  \equiv
 { \spa1.2^6  }
{ \spb{p_1}.{p_2} \over
\spa{p_1}.{p_2}\spa{p_1}.{1}\spa{p_1}.2\spa{p_2}.1 \spa{p_2}.2 }
\ee
The  $\hat M_3^{MHV}$
and $\hat M_4^{MHV}$ are not true amplitudes since their arguments are
not constrained by
momentum conservation. 
We have made  specific (natural) choices for $\hat M_3^{MHV}$
and $\hat M_4^{MHV}$. These coincide with the on-shell amplitudes
where $\bar\lambda_{1,2}$ are shifted as
in eq.~(\ref{SoftShiftEQ}).

Expression (\ref{TreeLiftingEQ4}) can be verified recursively by considering a BCFW shift~\cite{Britto:2005fq}
of the two negative helicity legs:
\be
\lambda_1\to \lambda_1+z\lambda_2, \qquad
\bar\lambda_2\to \bar\lambda_2-z\bar\lambda_1.
\ee
Under this shift the amplitude has poles when $\la 1 p_a\ra\to 0$, of
the form
\begin{align}
{\spb{{p_a}}.{1}\over \spa{\hat 1}.{p_a}}
{ \spa{1}.{2}^6   \over  \spa{2}.{p_a}^6  }
M_{n-1}^{MHV}(\hat K^-,\hat 2^-, \cdots [p_a]\cdots)
\label{Wtreepoles}
\end{align}
where $[p_a]$ denotes the absence of leg $p_a$ from the argument
list and 
\be
\hat K = \lambda_{p_a} { \la 2 | (k_1+p_a) |\over  \spa{2}.{p_a} } 
\; . 
\ee
 Applying the shift to  (\ref{TreeLiftingEQ4}) we find poles for
the same values of $z$.  The residues of these  poles receive
contributions from  terms where leg $p_a$ lies within $\hat M_4$ and
from terms where it lies within the soft lifting function. In the
latter case the residue can be expressed in terms of an $n-1$ point
version of  (\ref{TreeLiftingEQ4}), while for the former case we
use the $n-1$ point version of (\ref {SoftisTree}). Combining the two
we find precisely the required complex factorisation of the tree
amplitude (\ref{Wtreepoles}).  Since the expression matches the tree
amplitude for $n=3,4$ this guarantees it is correct for all $n$.
(The large $z$ behaviour also follows that of the  gravity tree
amplitude (\ref {SoftisTree}).)

Equations (\ref{TreeLiftingEQ}) 
and  (\ref{TreeLiftingEQ4})
look like the first two members of a chain of relationships of which
eq.~(\ref{InverseSoftEX}) 
looks like the last. This is particularly true if we sum over $p$ in
eq.(\ref{TreeLiftingEQ}) and divide by $(n-2)$.  However we have been
unable to construct other possible terms in such a sequence.

\section{Application: Rational terms in $\NeqFour$ one-loop amplitudes}

A one-loop graviton scattering amplitude can receive contributions
from a range of particle types circulating in the loop.  We denote the
contribution from a particle of spin-$s$ to the graviton scattering
amplitude by $M^{[s]}_n$ (with $M^{[0]}_n$ representing a real scalar).  
In a supergravity theory there can be contributions from  minimally  coupled matter
multiplets. 
The contributions to graviton scattering amplitudes from the various
supergravity multiplets are~\cite{GravityStringBasedB}
\begin{align}
 M_n^{\Neq8} =&   M_n^{[2]} +8 M_n^{[3/2]}+28 M_n^{[1]} +56 M_n^{[1/2]} +
 70 M_n^{[0]}
\notag\\
M_n^{\Neq6,matter} =&   M_n^{[3/2]}+6 M_n^{[1]} +15 M_n^{[1/2]} +
20 M_n^{[0]}
\notag\\
  M_n^{\Neq4} =&   M_n^{[2]} +4 M_n^{[3/2]}+6 M_n^{[1]} +4 M_n^{[1/2]} +
 2M_n^{[0]}
\notag
\\
  M_n^{\Neq4,matter} =&   M_n^{[1]} +4 M_n^{[1/2]}+ 6M_n^{[0]} 
\notag
\\
 M_n^{\Neq1,matter} =& M_n^{[1/2]} +
 2M_n^{[0]}
\end{align}

In terms of the supersymmetric matter contributions the one-loop $\NeqFour$ supergravity  amplitude is
\be
M^{\NeqFour}_n= M^{\NeqEight}_n- 4M^{\NeqSix,matter}_n   + 2 M^{\NeqFour,matter}_n
\ee
Extensions to the basic $\NeqFour$ theory can be obtained from
variants of this formula~\cite{Dunbar:1999nj} .  For example the case of ``type I'' theory
which is the dimensional reduction of $\NeqOne$ ten dimensional
supergravity is
\be
M^{\NeqFour*}_n= M^{\NeqEight}_n- 4M^{\NeqSix,matter}_n   + 8 M^{\NeqFour,matter}_n
\ee
 and if the supergravity is coupled to a $\NeqFour$ gauge theory with
 gauge group $G$, 
\be
M^{\NeqFour*,G}_n= M^{\NeqEight}_n- 4M^{\NeqSix,matter}_n   + (8+\dim G)  M^{\NeqFour,matter}_n
\ee

A general $n$-point one-loop amplitude in a massless theory such as
gravity or QCD can be expanded in terms of loop momentum integrals, $I_m[P^d(\ell)]$,
where $m$ denotes the number of vertices in the loop and $P^d(\ell)$ is
a polynomial of degree $d$ in the loop momentum $\ell$.  For gravity
we expect $d=2m$ since the three-point vertex is quadratic in
momentum. While  for supergravity theories the naive expectation would be
$d=2m-r$  where $r=8,6,4,2$ for $\NeqEight,6,4,1$ respectively~\cite{GravityStringBasedA,GravityStringBasedB} .  
However,  the evidence from explicit calculations suggests an
effective degree for loop momentum
polynomial~\cite{BjerrumBohr:2006yw,MaxCalcsB,Dunbar:2010fy,Bern:2007xj,BjerrumBohr:2008vc} of
\be
\deff=(m+4)-r 
\label{deffEQ}
\ee
with $r=4$ for $\NeqFour$, $r=7$ for $\NeqSix$ and $r=8$ for
$\NeqEight$. 
Performing a Passarino-Veltman~\cite{PassVelt}  reduction on the loop momentum
integrals yields an amplitude (to $O(\eps)$ in the dimensional
reduction parameter $\eps$), 
\be
 \Aloop_n=\sum_{i\in \cal C}\, c_i\, I_4^{i}
 +\sum_{j\in \cal D}\, d_{j}\, I_3^{j}
 +\sum_{k\in \cal E}\, e_{k} \,   I_2^{k}
+R_n\,,
\label{basisequn}
\ee
where $c_i,d_i,e_i$ and $R_n$ are rational functions and the $I_4$,
$I_3$, and $I_2$ are scalar box, triangle and bubble functions
respectively.  The mathematical form of these integral functions
depends on whether the momenta flowing into a vertex are null
(massless) or not (massive).    For amplitudes with the $\deff$ of
eq.~(\ref{deffEQ}),  the expansion of~(\ref{basisequn}) simplifies:
$\NeqEight$ amplitudes contain only box integral contributions ~\cite{BjerrumBohr:2006yw,MaxCalcsB},
$\NeqSix$ amplitudes  contain only box and triangle contributions, while  
for $\NeqFour$ since $\deff=m$ the amplitude has the full spectrum of integral functions and rational
terms.
While this coincides with the power counting for Yang-Mills, supersymmetry 
imposes other simplifications on the $\NeqFour$ supergravity amplitudes, in particular the vanishing 
of the ``all-plus'' and ``single-minus'' one-loop amplitudes which
simplifies the factorisation structure.
  
%The simplifications that appear at one-loop also have implications at multi-loop level: there is the conjecture of perturbative finiteness for
%$\NeqEight$ supergravity~\cite{Bern:2006kd} while for $\NeqFour$ the power count of eq.~(\ref{deffEQ}) suggests that the three-loop amplitudes may be 
%finite~\cite{Dunbar:2010fy}, as has recently  been explicitly calculated for the four-point amplitude~\cite{Bern:2012cd}.   

In terms of the expansion (\ref{basisequn}),  the $n$-graviton MHV amplitude for a $\NeqFour$ matter multiplet is~\cite{Dunbar:2010fy}
\begin{multline}
  M_n^{\NeqFour,matter} (1^-, 2^-, 3^+, \ldots, n^+) = \\
  \quad {(-1)^n\over 8} \, \spa1.2^8 \sum_{2 < a < b \leq n \atop 1\in
    M, 2 \in N} \left( { \spa{1}.{a}\spa{2}.{a}\spa{1}.{b}\spa{2}.{b}
      \over \spa{a}.b^2 \spa{1}.{2}^2 } \right)^2 h(a, M, b) h(b, N,
  a)
  \tr^2[a\, M\, b\, N]\, I_4^{aMbN, \trunc} \ \\
  \quad\quad+\sum_{1\in A,2\in B} e_{A;B} I_2 (K_A^2) +R_n,\hfill
  \label{MHVboxes}\end{multline}
The summation over boxes is over subsets $M,N$ such that $1\in M$,
$2\in N$ and
$M\cup N\cup\{a,b\}=\{1,\cdots n\}$. The summation over bubbles is
over subsets $A$ and $B$ where  $1\in A$, $2\in B$,  each contain at least one positive helicity
leg and $A\cup B=\{1,\cdots n\}$.   The truncated box-functions are
the specific combinations of the
scalar box and triangle functions
\be
 I_4^{i,\trunc} 
=I_4^{i} +\sum_{j}  \tilde b_{ij} I_3^{j}
\ee
which are IR and UV finite (see, for example,  the appendix of
ref.~\cite{Dunbar:2011xw} for explicit expressions). Using truncated
box functions automatically implements the constraints from IR and UV
singularities~\cite{Dunbar:1995ed}  with the single remaining constraint
$\sum e_{A;B}=0$.

\begin{figure}[H]
  \begin{center}
    {
      \begin{picture}(150,144)(0,0)
    \SetOffset(50,72)
        \Line(-40,0)(0,40)
        \Line(0,40)(40,0)
        \Line(40,0)(0,-40)
        \Line(0,-40)(-40,0)
        \Line(0,40)(0,60)
        \Line(0,-40)(0,-60)
        \Text(0,62)[bc]{$b^+$}
        \Text(0,-62)[tc]{$a^+$}
        \Line(40,0)(60,0) \Line(40,0)(55,15) \Line(40,0)(55,-15)
        \Line(-40,0)(-60,0) \Line(-40,0)(-55,15) \Line(-40,0)(-55,-15)
        \Text(62,1)[lc]{$2^-$}
        \Text(-62,1)[rc]{$1^-$}
        \Vertex(53,8){0.5} \Vertex(54,4.5){0.5}
        \Vertex(53,-8){0.5} \Vertex(54,-4.5){0.5}
        \Vertex(-53,8){0.5} \Vertex(-54,4.5){0.5}
        \Vertex(-53,-8){0.5} \Vertex(-54,-4.5){0.5}
        \Text(-75,0)[rc]{$\displaystyle M \left\{\vphantom{\frac12}\right.$}
        \Text(75,0)[lc]{$\displaystyle
          \left.\vphantom{\frac12}\right\}N$}
      \end{picture}
     \begin{picture}(150,144)(0,0)
        \SetOffset(125,72)
 %       \Line(-40,0)(0,40)
 %       \Line(0,40)(40,0)
 %       \Line(40,0)(0,-40)
 %       \Line(0,-40)(-40,0)
 %       \Line(0,40)(0,60)
 %       \Line(0,-40)(0,-60)
 %       \Text(0,62)[bc]{$b^+$}
 %       \Text(0,-62)[tc]{$a^+$}
\BCirc(0,0){40}
        \Line(40,0)(60,0) \Line(40,0)(55,15) \Line(40,0)(55,-15)
        \Line(-40,0)(-60,0) \Line(-40,0)(-55,15) \Line(-40,0)(-55,-15)
        \Text(62,1)[lc]{$2^-$}
        \Text(-62,1)[rc]{$1^-$}
        \Vertex(53,8){0.5} \Vertex(54,4.5){0.5}
        \Vertex(53,-8){0.5} \Vertex(54,-4.5){0.5}
        \Vertex(-53,8){0.5} \Vertex(-54,4.5){0.5}
        \Vertex(-53,-8){0.5} \Vertex(-54,-4.5){0.5}
        \Text(-75,0)[rc]{$\displaystyle A \left\{\vphantom{\frac12}\right.$}
        \Text(75,0)[lc]{$\displaystyle
          \left.\vphantom{\frac12}\right\}B$}
      \end{picture}
    }
    \\
    \caption{\captionfigA \label{BoxFigure} }
  \end{center}
\end{figure}

\def\TEST{
\begin{figure}[t]
  \centerline{\includegraphics*[width=8.5cm]{fig1.eps}}
 \caption{
 \captionfigA.} 
 \label{BoxFigure}
 \end{figure}
} %TEST

The coefficients of the scalar bubbles, $e_{A;B}$, are presented
explicitly in ref.\cite{Dunbar:2011xw} using canonical
forms~\cite{Dunbar:2009ax}. The precise form of these does not impact
on the rational terms (unlike the box coefficients) so we do not
present them here. 

The $n$-point rational term, which completes the $n$-point amplitude
$M(1^-,2^-,3^+,\cdots n^+)$ was proposed in ref.~\cite{Dunbar:2011dw}
and shown numerically to have the correct soft and collinear limits
for $n \leq 10$,
\be
R_n 
=(-1)^n{  \spa{1}.{2}^4}  \biggl( {R_n^0\over 2}+   \sum_{r=3}^{n-2} R_n^r  \biggr) 
\ee
In the above
\be
R_n^0 = \sum_{a,b \in P^+}  R_n^{0;a,b}
\ee
where
\begin{align}
R_n^{0;a,b} = %\sum_{box-configurations} 
\sum_{1\in M, 2\in N} {\spb{a}.b^2 \over \spa{a}.b^2 }& h(a,M,b)h(b,N,a) 
%\ee
%\begin{equation}
\times
%\end{equation}
\left(\spa{1}.a \spa{2}.a \spa{1}.b \spa{2}.b
\right)^2 
\end{align}
and there is a contribution for each box integral function present in the amplitude.
The $R_n^{0;a,b}$ contain spurious quadratic singularities,
$\spa{a}.b^{-2}$, which are
necessary to cancel those in the box integral
contributions~\cite{Dunbar:2011xw} and 
take the form
\be
c_{box}  \times   { s_{ab}^2 \over  2\tr(aMbN)^2 }
\ee
Note that this $R_n^{0;a,b}$-term taken together with the
corresponding box integral contribution  has no
phase-singularity as $a$ and $b$ become collinear.

The remaining $R_n^r$
are 
\be 
R_n^r = \sum_{P^r\subset P^+,|P^r|=r}    C_{r} [P^r]  
\times \SoftP^{n-2-r}[P^+-P^r ] \;\;\; , \;\;  r=3,\cdots n-2
\ee
The sum is over all subsets $P^r$ of $P^+$ of
length $r$ (of which there are
$(n-2)!/r!/(n-2-r)!$) and  $P^+-P^r$ are the remaining positive helicity legs.
The $C_r$ are 
\be
C_r[ P^r]\equiv
\sum_{perms}  
{ \spb{p_1}.{p_2} \spb{p_2}.{p_3} \cdots \spb{p_r}.{p_1}
\over
\spa{p_1}.{p_2} \spa{p_2}.{p_3} \cdots \spa{p_r}.{p_1} }
\ee
where the sum over permutations is over the $(r-1)!/2$ cyclically independent
choices of orderings of $\{p_1,\cdots p_r\}$ (we take the cycle
$(p_1,p_2,\ldots,p_{n-1},p_n)$ to be equivalent to the complete reversal
$(p_n,p_{n-1},\ldots,p_{2},p_1)$).  This implies the definition 
\be
C_2[\{p_1,p_2\}]=\frac{1}{2} { \spb{p_1}.{p_2}^2 \over \spa{p_1}.{p_2}^2 }
\ee
For example, 
the five~point~\cite{Dunbar:2010fy}  and six-point expressions are 
\begin{align}
R_5&={-\spa1.2^4}   \biggl(  {R_5^0 \over 2}+  {\spb3.4\spb4.5\spb5.3
  \over \spa3.4\spa4.5\spa5.3 } \biggr)
\notag\\
R_6 &={\spa1.2^4} \biggl(  { R_6^0\over 2}
+\biggl(
 {\spb3.4\spb4.5\spb5.3 \over \spa3.4\spa4.5\spa5.3 }
\sum_{j\neq 6}  {\spb6.j \spa1.j\spa2.j \over \spa6.j \spa1.6\spa2.6 } 
+ \{ 6\leftrightarrow 3,4,5 \} \biggr) 
\notag\\
& +\biggl( 
{\spb3.4\spb4.5\spb5.6\spb6.3  \over \spa3.4\spa4.5\spa5.6\spa6.3 }
+
{\spb3.4\spb4.6\spa6.5\spb5.3  \over \spa3.4\spa4.6\spa6.5\spa5.3 }
+
{\spb3.5\spb5.4\spb4.6\spb6.3  \over \spa3.5\spa5.4\spa4.6\spa6.3 }
 \biggr) \biggr)
\label{EQsmalln}
\end{align}

We will show analytically that $R_n$ has the correct soft and
collinear limits. While our expression for $R_n$ is conjectural  for $n> 5$,
experience strongly suggests that expressions with the correct soft and
collinear factorisations are likely to be correct. 
An important step, which we require to establish the collinear limits
of $R_n$, is the identification
\be
{ R^0_n \over 2}   =  \sum_{\{p_1,p_2\} \subset P^+}   C_2[\{p_1,p_2\}]\times  \SoftP^{n-4} [ \{
p_1,p_2\} ;  P^+-\{ p_1,p_2\}  ]
\label{EQR0identity}
\ee
which we would naturally label as $R_n^2$.   To do so we use an identity for
the quadratic product of half-soft functions,
\be
\sum_M  h( a, M+c,b) h(b,N+d,a) = \sum_M h( c, M+a,d) h(d,N+b,c)  
\ee
where the summation over $M$ is over all  subsets of
$\{1,2,\ldots,n\}-\{a,b,c,d\}$ and
$N=\{1,2,\ldots,n\}-\{a,b,c,d\}-M$.  We have verified this identity at
specific kinematic points for $n\leq 12$. Using this identity and the definition of the soft-lifting
function~eq.(\ref{SoftDefineEQ}) we have
\begin{align}
R_n^{0;p_1,p_2} = %\sum_{box-configurations} 
& {\spb{p_1}.{p_2}^2 \over \spa{p_1}.{p_2}^2 } \sum_{ M} h(p_1,M+1,p_2)h(p_2,N+2,p_1) 
%\ee
%\begin{equation}
\times
%\end{equation}
\left(\spa{1}.{p_1} \spa{2}.{p_1} \spa{1}.{p_2} \spa{2}.{p_2}
\right)^2 
\notag\\
=& 
 {\spb{p_1}.{p_2}^2 \over \spa{p_1}.{p_2}^2 }  \sum_{M} h(1,M+p_1,2)h(2,N+p_2,1) 
%\ee
%\begin{equation}
\times
%\end{equation}
\left(\spa{1}.{p_1} \spa{2}.{p_1} \spa{1}.{p_2} \spa{2}.{p_2}
\right)^2
\notag\\
=& 
 {\spb{p_1}.{p_2}^2 \over \spa{p_1}.{p_2}^2 } \SoftP^{n-4} [ \{
p_1,p_2\} ;  P^+-\{ p_1,p_2\}  ]=
2C_2[\{ p_1,p_2\}]\times 
\SoftP^{n-4} [ \{
p_1,p_2\} ;  P^+-\{ p_1,p_2\}  ]
%\ee
%\begin{equation}
\end{align}
Consequently, we can rewrite $R_n$ in the unified form
\be
R_n 
=(-1)^n{  \spa{1}.{2}^4}   \sum_{r=2}^{n-2} R_n^r  
\label{EQalln}\ee

%The expression (\ref{EQalln}) has been derived by taking known results
%and using soft-lifting functions to generate all-$n$ expression.  

We now make the observation that this expression for the rational term $R_n$, remarkably, also has an
algebraic diagrammatic  expression akin to that for the MHV tree
amplitude~\cite{Nguyen:2009jk}.  
First, consider the cycle term $C_r$,
\be
C_r[\{p_1,\ldots,p_r\}] = \sum_{permutations}    { \spb{p_1}.{p_2} \spb{p_2}.{p_3} \cdots \spb{p_{r}}.{p_1}
\over
\spa{p_1}.{p_2} \spa{p_2}.{p_3} \cdots \spa{p_{r}}.{p_1} }
\ee
where the sum is over all permutations of the $r$ positive helicity
legs in the cycle.  Each term in the sum may be interpreted as a one-loop link 
graph of the type shown in fig.~\ref{CycleFigure} where all $r$ positive
helicity legs lie in the loop. 
In general $R_n^r$ contains $C_r$ factors multiplied by soft-lifting
functions.  Previously we saw how  the soft-lifting factors could be expressed
in terms of link tree diagrams emanating from seed points
(\ref{twigs}). This motivates expressing  $R_n$  as
\be
R_n^{MHV} 
=(-1)^n {\spa1.2^4 }
\sum_{\rm one-loop} \left(
\prod_{\rm edges\; :ab} {
\spb{a}.b \over \spa{a}.b }
\right)
\left(
\prod_{\rm vertices\; :a}  
( \spa{a}.1\spa{a}.2 )^{deg(a)-2} 
\right)
\label{DiagramLoopEQ}
\ee
where the sum is over all distinct, connected, one-loop link graphs
involving vertices labelled by the $n-2$ positive helicity legs.  
Terms within  $R_n^r$ correspond to graphs with $r$ vertices
in the loop.

\begin{figure}[H]
  \begin{center}
    {
      \begin{picture}(100,100)(0,0)
    \SetOffset(50,50)
\SetWidth{2}
%\BCirc(0,0){40}
\CArc(0,0)(40,180,90)
\DashCArc(0,0)(40,90,180){3}
\SetWidth{1}
\BCirc(40,0){5}
\BCirc(-40,0){5}
\BCirc(0,40){5}
\BCirc(0,-40){5}
\BCirc(28.28,28.28){5} 
%\BCirc(-28.28,28.28){5} 
\BCirc(28.28,-28.28){5} 
\BCirc(-28.28,-28.28){5} 
      \end{picture}
     }
    \\
    \caption{\captionfigD \label{CycleFigure} }
  \end{center}
\end{figure}
The connection between  fig.~\ref{CycleFigure}  and the corresponding term in $C_r[P^r]$ is
fairly clear when we note that $\rm deg(a)=2$ for each vertex and the sum
over permutations is simply the sum over
diagrams.  In the $R_n^r$ term the $C_r$ is multiplied by
$\SoftP^{n-2-r}[P^+-P^r]$.   The individual terms in the soft-lifting
function correspond to individual diagrams of the type of
fig.~\ref{FigO}.  Multiplying the two factors, individual terms will
correspond exactly to diagrams where the trees of fig.~\ref{FigO}  are attached to the
loop of fig.~\ref{CycleFigure} as in fig.~\ref{CycleFigureB}.
\begin{figure}[H]
  \begin{center}
    {
      \begin{picture}(150,140)(0,0)
    \SetOffset(75,60)
\SetWidth{2}
\CArc(0,0)(40,180,90)
\DashCArc(0,0)(40,90,180){3}
\Line(40,0)(80,0)
\Line(80,0)(120,0)
\Line(40,0)(70,25)
\Line(40,0)(70,-25)
\Line(0,40)(0,80)

\Line(-40,0)(-80,0)
\Line(-80,0)(-110,25)
\Line(-80,0)(-110,-25)
\SetWidth{1}
\BCirc(40,0){8}
\BCirc(-40,0){8}
\BCirc(0,40){8}
\BCirc(0,-40){8}
\BCirc(28.28,28.28){8} 
%\BCirc(-28.28,28.28){10} 
\BCirc(28.28,-28.28){8} 
\BCirc(-28.28,-28.28){8} 
\BCirc(80,0){8}
\BCirc(120,0){8}
\BCirc(70,25){8}
\BCirc(70,-25){8}
%\Text(80,0)[]{$q$}
\Text(-40,0)[]{$p_1$}
\Text(0,40)[]{$p_2$}
\Text(40,0)[]{$p_3$}
\BCirc(-80,0){8}
\BCirc (-110,25){8}
\BCirc (-110,-25){8}
\BCirc(0,80){8}
      \end{picture}
     }
    \\
    \caption{\captionfigE \label{CycleFigureB} }
  \end{center}
\end{figure}

We now present a diagrammatic proof that the rational terms have the
correct soft and collinear limits.  For the $\NeqFour$ MHV amplitude
several of these vanish since the one-loop amplitude with a single
negative helicity leg vanishes in a supersymmetric theory. 
Consequently, there is no soft-singularity when one of the negative
helicity legs vanishes since the target amplitude vanishes.
Similarly, in the $m^-b^+$ collinear limit there is no  $S^{-+}_-$
term.  In the $a^+b^+$  collinear limit only the
$S^{++}_-$ splitting function is non-vanishing.
Furthermore  the MHV amplitude has no multi-particle poles for real
momenta.   Although the power-counting of $\NeqFour$ supergravity is
the same as Yang-Mills, the lack of amplitudes with a single negative
helicity leg gives the MHV amplitude a simpler factorisation structure. 

The soft and collinear limit constraints  apply to whole amplitudes, but for our amplitude
the integral function contributions and the rational terms  have most of the appropriate limits
independently. The exception to this is in the $a^+b^+$ limit where
the $C_2[\{a,b\}]$ terms combine with the box integral contributions~\cite{Dunbar:2011xw}.

\subsection{Diagrammatic Proof:  $a^+b^+$-Collinear Limit}
We can use the diagrammatic representation to give a proof that $R_n$ 
has the correct   $a^+,b^+$ collinear limit.   Consider the limit of the $n$-point amplitude when $k_a \longrightarrow z K$, $k_b \longrightarrow
\bar z K$, with $\bar z =1-z$.  
Consider a diagram contributing to  the $n-1$ point amplitude and focus on the
vertex labelled $K$ as in fig.~\ref{CycleFigureK}. This vertex may, or may not lie within a loop.  
\begin{figure}[H]
  \begin{center}
    {
      \begin{picture}(250,140)(0,0)
    \SetOffset(100,70)
\SetWidth{2}
\Line(0,0)(40,0)
\DashLine(40,0)(80,0){3}
\Line(0,0)(-40,0)
\DashLine(-40,0)(-80,0){3}
\Line(0,0)(-28.28,-28.28)
\DashLine(-28.28,-28.28)(-56.56,-56.56){3}
\Line(0,0)(-28.28,28.28)
\DashLine(-28.28,28.28)(-56.56,56.56){3}
\Line(0,0)(28.28,-28.28)
\DashLine(28.28,-28.28)(56.56,-56.56){3}
\Line(0,0)(28.28,28.28)
\DashLine(28.28,28.28)(56.56,56.56){3}
\BCirc(0,0){8}
\SetWidth{1}
\Text(0,0)[]{$K$}
\end{picture}   
      }
    \caption{\captionfigK\label{CycleFigureK} }
  \end{center}
\end{figure}
In general this vertex 
will have $n_K>0$ legs 
attached to it and will have an associated vertex factor
\be
(\spa{1}.K\spa2.K)^{n_K-2}
\ee
The $n$-point diagrams which contribute to this vertex in the
collinear limit must have legs $a$ and $b$ connected by a single link.
These 
are of the form show in fig.~\ref{CycleFigureG}. The link
contributes
\be
{\spb{a}.b\over \spa{a}.b }  \coli{a}{b}  -z\bar z S^{++}_-
\ee
with the other links smoothly going to those involving $K$ in fig.~\ref{CycleFigureK}.
\begin{figure}[H]
  \begin{center}
    {
      \begin{picture}(250,140)(0,0)
    \SetOffset(100,70)
\SetWidth{2}
\Line(30,0)(60,0)
\DashLine(60,0)(90,0){3}
\Line(-30,0)(-60,0)
\DashLine(-60,0)(-90,0){3}
\Line(-30,0)(-58.28,-28.28)
\DashLine(-58.28,-28.28)(-86.56,-56.56){3}
\Line(-30,0)(-58.28,28.28)
\DashLine(-58.28,28.28)(-86.56,56.56){3}
\Line(30,0)(58.28,-28.28)
\DashLine(58.28,-28.28)(86.56,-56.56){3}
\Line(30,0)(58.28,28.28)
\DashLine(58.28,28.28)(86.56,56.56){3}
\SetColor{Red}
\Line(-30,0)(30,0)
\BCirc(-30,0){8}
\BCirc(30,0){8}
\SetColor{Black}
\SetWidth{1}
\Text(-30,0)[]{$a$}
\Text(30,0)[]{$b$}

\end{picture}   
      }
    \caption{\captionfigL\label{CycleFigureG} }
  \end{center}
\end{figure}
In these diagrams the same set of initial legs must be attached to
vertices $a$ and $b$. All possible combinations are present. In
general $n_a$ of the original legs will be attached to vertex $a$ and $n_b$
to vertex $b$ with $n_a+n_b=n_K$.  There will be $n_k!/n_a!/(n_K-n_a)!$
independent choices of which legs are attached.  In the collinear limit the
factors associated with the two vertices give
\be
\Bigl(\spa{1}.a\spa2.a\Bigr)^{n_a-1}
\Bigl(\spa{1}.b\spa2.b\Bigr)^{n_b-1}
\coli{a}{b}  
z^{n_a-1} \bar z^{n_b-1}  \Bigl(\spa{1}.K\spa2.K\Bigr)^{n_a+n_b-2}
\ee
Summing over all contributions gives the descendant
diagram times a factor of
\be
\sum_{n_a=0}^{n_K}
{ n_K! \over n_a! (n_K-n_a)! }  z^{n_a}\bar z^{n_K-n_a} \times -S^{++}_-
=(z+\bar z)^{n_K}  \times -S^{++}_-
=-S^{++}_-
\ee
Note that this proof relies upon the fact that the loop-diagrams shown in
fig.~\ref{CycleFigureM} do not contribute to the $a^+b^+$ collinear limit
but instead cancel against the box integral
contributions in this limit~\cite{Dunbar:2011xw}.  

\begin{figure}[H]
  \begin{center}
    {
      \begin{picture}(250,140)(0,0)
    \SetOffset(100,70)
\SetWidth{2}
\Line(30,0)(60,0)
\DashLine(60,0)(90,0){3}
\Line(-30,0)(-60,0)
\DashLine(-60,0)(-90,0){3}
\Line(-30,0)(-58.28,-28.28)
\DashLine(-58.28,-28.28)(-86.56,-56.56){3}
\Line(-30,0)(-58.28,28.28)
\DashLine(-58.28,28.28)(-86.56,56.56){3}
\Line(30,0)(58.28,-28.28)
\DashLine(58.28,-28.28)(86.56,-56.56){3}
\Line(30,0)(58.28,28.28)
\DashLine(58.28,28.28)(86.56,56.56){3}
\SetColor{Red}
%\Line(-30,0)(30,0)
\CArc(0,-30)(42.42,45,135)
\CArc(0, 30)(42.42,225,315)
\BCirc(-30,0){8}
\BCirc(30,0){8}
\SetColor{Black}
\SetWidth{1}
\Text(-30,0)[]{$a$}
\Text(30,0)[]{$b$}

\end{picture}   
      }
    \caption{\captionfigM\label{CycleFigureM} }
  \end{center}
\end{figure}

\subsection{Diagrammatic Proof:  Soft Limit}

We can also use the diagrammatic representation to show that $R_n$ has
the correct soft limit as $k_n\longrightarrow 0$. 
Consider a link-diagram of the $n-1$ point amplitude and examine the
set of $n$-point diagrams which might give this in the soft limit.  In eq.~(\ref{DiagramLoopEQ})
there is no soft-singularity from the links and the only possible
singularity is from the vertex contribution when $\deg(n)=1$.  These
factors arise  where a link from vertex $n$ is added to the $n-1$ point
diagram  as in figure~\ref{CycleFigureN}.

 \begin{figure}[H]
  \begin{center}
    {
  \begin{picture}(110,80)(0,0)
    \SetOffset(10,50)
\SetWidth{2}
\CArc(0,0)(30,180,90)
\DashCArc(0,0)(30,90,180){3}
\Line(30,0)(60,0)
\SetWidth{1}
\BCirc(30,0){5}
\BCirc(-30,0){5}
\BCirc(0,30){5}
\BCirc(0,-30){5}
\BCirc(21.21,21.21){5} 
\BCirc(21.21,-21.21){5} 
\BCirc(60,0){5}
\SetColor{Red}
\Line(-21.21,-21.21)(-40,-30)
\BCirc(-40,-30){5}
\SetColor{Black}
\BCirc(-21.21,-21.21){5} 
\Text(-40,-30)[]{\small $n$}
      \end{picture}
\begin{picture}(110,80)(0,0)
    \SetOffset(10,50)
\SetWidth{2}
\CArc(0,0)(30,180,90)
\DashCArc(0,0)(30,90,180){3}
\Line(30,0)(60,0)
\SetWidth{1}
\BCirc(30,0){5}
\BCirc(-30,0){5}
\BCirc(0,30){5}
\BCirc(0,-30){5}
\BCirc(21.21,21.21){5} 
\BCirc(21.21,-21.21){5} 
\BCirc(-21.21,-21.21){5} 
\BCirc(60,0){5}
\SetColor{Red}
\Line(30.0,0)(50,-10)
\BCirc(50,-10){5}
\SetColor{Black}
\BCirc(30.0,0.0){5} 
\Text(50,-10)[]{\small $n$}
      \end{picture}
\begin{picture}(110,80)(0,0)
    \SetOffset(10,50)
\SetWidth{2}
\CArc(0,0)(30,180,90)
\DashCArc(0,0)(30,90,180){3}
\Line(30,0)(60,0)
\SetWidth{1}
\BCirc(30,0){5}
\BCirc(-30,0){5}
\BCirc(0,30){5}
\BCirc(0,-30){5}
\BCirc(21.21,21.21){5} 
\BCirc(21.21,-21.21){5} 
\BCirc(-21.21,-21.21){5} 
\BCirc(60,0){5}
\SetColor{Red}
\Line(60.0,0)(60,-20)
\BCirc(60,-20){5}
\SetColor{Black}
\BCirc(60.0,0.0){5} 
\Text(60,-20)[]{\small $n$}
      \end{picture}
\begin{picture}(100,80)(0,0)
    \SetOffset(40,50)
\SetWidth{2}
\Text(-50,0)[]{$\longrightarrow$}
\CArc(0,0)(30,180,90)
\DashCArc(0,0)(30,90,180){3}
\Line(30,0)(60,0)
\SetWidth{1}
\BCirc(30,0){5}
\BCirc(-30,0){5}
\BCirc(0,30){5}
\BCirc(0,-30){5}
\BCirc(21.21,21.21){5} 
\BCirc(21.21,-21.21){5} 
\BCirc(-21.21,-21.21){5} 
\BCirc(60,0){5}
      \end{picture}
      }
    \caption{\captionfigN\label{CycleFigureN} }
  \end{center}
\end{figure}

In the soft limit, the diagram where leg $n$ is attached to vertex $j$ gives a
contribution of
\be
-{  \spb{j}.n \spa{1}.j  \spa{2}.j \over \spa{j}.n  \spa{1}.n
  \spa{2}.n  } \times (n-1) \hbox{\rm -point diagram}
\ee
Summing the different contributions gives  the correct soft
factor of~eq.(\ref{SoftFactorEq}) with $a,b=1,2$.  

\subsection{Diagrammatic Proof:  $m_1^-n^+$-Collinear Limit}

Finally we consider  the limit of the $n$-point amplitude when
$k_{m_1} \longrightarrow z K$, 
$k_n \longrightarrow
\bar z K$, with $\bar z =1-z$.  
Examining, eq.~(\ref{DiagramLoopEQ}), we see that contributions only
arise in this limit when leg $n$ is an isolated leg, much  as the soft-leg is isolated
in fig.~\ref{CycleFigureN}.  Each  diagram gives a
$\spa{m_1}.{n}^{-1}$ singularity but
summing over the diagrams reduces this to a collinear phase singularity as
required. 
The diagram where vertex $n$ is attached to vertex $j$
gives the descendant diagram times a factor of
\be
{ \spb{n}.j \spa{m_1}.j \spa{m_2}.j 
\over 
\spa{n}.j \spa{m_1}.n \spa{m_2}.n }
\ee
Using the Schouten identity, 
\be
{\spa{m_1}.j \over \spa{n}.j }
={ \spa{m_1}.X \over \spa{n}.X } +{\spa{m_1}.n \spa{j}.X \over
  \spa{n}.j \spa{n}.X } 
\ee
The second term cancels the singularity and these
terms do not contribute. Summing over the remaining contribution from
each diagram we have 
\begin{align}
\sum_{j\in P^+-n} { \spb{n}.j \spa{m_1}.X \spa{m_2}.j 
\over 
\spa{n}.X\spa{m_1}.n \spa{m_2}.n }
=-{   \spa{m_1}.X \over \spa{n}.X\spa{m_1}.n \spa{m_2}.n  }   \sum_{j\in
  P^+-n} { \spb{n}.j  \spa{j}.{m_2} 
}
\notag\\
={   \spa{m_1}.X \over \spa{n}.X\spa{m_1}.n \spa{m_2}.n  }
\sum_{j=m_1,m_2,n} { \spb{n}.j  \spa{j}.{m_2} }
\notag\\
={   \spa{m_1}.X \over \spa{n}.X\spa{m_1}.n \spa{m_2}.n  }  { \spb{n}.{m_1} \spa{m_1}.{m_2} }
\end{align}
Now in the collinear limit we can see this has the collinear phase singularity
\be
\coli{m_1}{p}
{z\over (1-z) }{\spb{n}.{m_1} \over \spa{n}.{m_1} }   =- {1\over z^2 } S^{-+}_+
\ee
Consequently, after including the pre-factor $\spa{m_1}.{m_2}^4$,
$R_n^r$ reduces to  $R_{n-1}^r$ times a soft factor  in the collinear
limit,
\be
(-1)^n \spa{m_1}.{m_2}^4 R_n^r  
\longrightarrow 
S^{-+}_+ \times 
(-1)^{n-1} \spa{K}.{m_2}^4 R_{n-1}^r \; . 
\ee

This completes the proof that the expression for the rational term
satisfies all real factorisation constraints.  Although this is short
of a full proof, experience strongly suggests that the expression is
correct.  Explicit computations using string based rules confirm the
expression for $n=4,5$~\cite{GravityStringBasedB,Dunbar:2010fy}. 
Confirmation of the result beyond $n=5$ will probably require
development of complex factorisation techniques~\cite{Dunbar:2010xk}
or use of the gravity-gauge relations~\cite{Bern:2010yg,Bern:2011rj}.

\section{Conclusions}

We have shown how soft-lifting functions can be used to generate
$n$-point MHV tree amplitudes and the rational pieces of the one-loop MHV
$n$-graviton amplitude in $\NeqFour$ supergravity.
In the latter case we have explicitly checked that our ansatz has the correct behaviour in
both the soft and collinear
limits. In some ways this is implementing the ideas of  the
``inverse-soft'' computations but we have applied it to one-loop
computation and used real-soft factorisation functions. In this
context soft-factorisation is not enough to generate the $n$-point
term and we must also appeal to  collinear factorisation properties.

Using the same functional rules as for the tree expression
of~\cite{Nguyen:2009jk}, our analytic expression is in one-one
correspondence with the set of one-loop connected
link diagrams. The  soft-lifting functions themselves have a diagrammatic
representation akin to that for the
MHV tree amplitudes.

In
the language of the soft-lifting functions we might think of lifting
some other 
low-point "seed" expression to obtain the corresponding $n$-point contribution. 
For example, 
the existence of seeds for next-to-MHV $\NeqFour$ and $\NeqOne$ MHV
amplitudes would provide fascinating generalisations of this process.

 The one-loop result is quite remarkable: although we might
expect tree relations to extend to the {\it integrands} of loop
expressions, there is no reason to expect them to survive the integration.
Indeed, it is not evident why the one-loop
link diagrams should give the rational terms of the $\NeqFour$ theory
rather than other supergravity theories.  The original tree diagrams
had an understanding in twistor space~\cite{ArkaniHamed:2009si} but it
is opaque as to why this would extend to $\NeqFour$ one-loop amplitudes.
It would be interesting 
 to consider $\NeqFour$ MHV amplitudes beyond 
one-loop however despite spectacular recent
progress~\cite{Bern:2012cd,BoucherVeronneau:2011qv}  there are very
few explicit supergravity calculations beyond one-loop.

\end{document}